\documentclass[runningheads]{llncs}
\usepackage[T1]{fontenc}
\usepackage{graphicx}
\usepackage{lscape}
\usepackage[para,online,flushleft]{threeparttable}
\usepackage{comment}
\usepackage{todonotes}

\usepackage{listings}
\definecolor{dkgreen}{rgb}{0,0.6,0}
\definecolor{darkred}{rgb}{0.3,0.1,0.1}
\definecolor{gray}{rgb}{0.5,0.5,0.5}
\definecolor{mauve}{rgb}{0.58,0,0.82}

\lstdefinestyle{DerekStyle}{ %
  language=C++,
  basicstyle=\ttfamily\ttfamily,
  numbers=left,
  numberstyle=\scriptsize\color{gray},  
  numbersep=5pt,                  
  backgroundcolor=\color{white},
  showspaces=false,               
  showstringspaces=false,         
  showtabs=false,
  frame=single,                   
  rulecolor=\color{black},
  tabsize=2,                      
  captionpos=b,                   
  breaklines=true,                
  breakatwhitespace=false,
  keywordstyle=\color{blue},          
  commentstyle=\color{dkgreen},       
  stringstyle=\color{mauve},         
  escapeinside={\%*}{*)},
  xleftmargin=4.0ex,
  morekeywords={for,each,between,can,reach,in,is,Sort,Print,From,size_t},
}   
\begin{document}
\title{Understanding GPU Triggering APIs \\for MPI+X Communication}
%
%
\author{Patrick G.\@ Bridges\inst{1}\orcidID{0000-0003-4801-03903}
\and
Anthony Skjellum\inst{2}\orcidID{0000-0001-5252-6600}
\and
Evan D.\hbox{} Suggs \inst{2}\orcidID{0000-0002-8210-8992}
\\ \and
Derek Schafer\inst{1}\orcidID{0000-0001-8438-5144}
\and
Purushotham V.\@ Bangalore\inst{3}\orcidID{0000-0002-1098-9997}
}
%
\authorrunning{P.\hbox{} Bridges et al.}
%
\institute{University of New Mexico, Albuquerque NM, 87131 \\
\email{\{patrickb,dschafer1\}@unm.edu}
\and
Tennessee Technological University, Cookeville, TN 38505 \\
\email{\{askjellum,esuggs\}@tntech.edu}
\and
The University of Alabama, Tuscaloosa, AL 35401 \\
\email{pvbangalore@ua.edu}
}

\maketitle              
\begin{abstract}
GPU-enhanced architectures are now dominant in HPC systems, but message-passing communication involving GPUs with MPI has proven to be both complex and expensive, motivating new approaches that lower such costs.
We compare and contrast stream/graph-, kernel-triggered, and GPU-initiated MPI communication abstractions, whose principal purpose is to enhance the performance of communication when GPU kernels create or consume data for transfer through MPI operations. Researchers and practitioners have proposed multiple potential APIs for 
GPU-involved communication
that span various GPU architectures and approaches, including MPI-4 partitioned point-to-point communication, stream communicators, and explicit MPI stream/queue objects. 
%
Designs breaking backward compatibility with MPI are duly noted. Some of these strengthen or weaken the semantics of MPI operations.
%
%
%
A key contribution of this paper is to promote community convergence toward 
common abstractions for GPU-involved communication by highlighting the common and differing goals and contributions of existing abstractions.
We describe the design space in which these abstractions reside, their implicit or explicit use of stream and other non-MPI abstractions, their relationship to partitioned and persistent operations, and discuss their potential for added performance, how usable these abstractions are, and where 
functional and/or semantic 
gaps exist.
%
%
Finally, we provide a taxonomy for these 
abstractions, including disambiguation of similar semantic terms, and consider directions for future standardization in MPI-5.

\end{abstract}

\section{Introduction}

Most MPI implementations now include support for GPU data paths, enabling standard MPI host-based APIs to move data to and from GPU memory. This feature is generally termed \emph{GPU-aware MPI} and significant progress has been made toward standardizing this implementation feature through work on \emph{MPI memory types}~\cite{mpi41}. Importantly, the MPI operation stages of \emph{initialization}, \emph{starting}, \emph{completion}, and \emph{freeing}
~\cite{mpi40}\footnote{The term `initiating' refers to a sequence of the initialization and starting phases.}
are all conducted by code running on the host CPU in this approach; GPU code that needs to cause or coordinate MPI operation state changes must do so indirectly by synchronizing with CPU code that makes those updates on behalf of the GPU code.

Integrating MPI into GPU \emph{control paths} is also important because it could eliminate the GPU/host communication latencies in current MPI implementations. This is a more challenging problem since it requires addressing multiple complex issues, including:
\begin{itemize}
    \item Which GPU control paths (stream execution and kernel execution) should support interacting with MPI operations?
    \item To what degree can current MPI abstractions be reused on GPU control paths? 
    \item What are the concurrency semantics of MPI operations on host and GPU control paths?
\end{itemize}

The myriad proposed approaches for addressing these issues~\cite{dinan_james_mpi-acx_2023,DBLP:conf/hpcasia/ElisPBB024,hewlett-packard_enterprise_mpix_start_2024,koziol_composable_2023,mpi40,namashivayam_exploring_2022,namashivayam_exploring_2023,venkatesh_mpi-gds_2017,zhou_mpix_2022} have focused on \emph{GPU-triggered communication APIs} in which host calls initialize and free MPI communication requests and the GPU control path starts
and in some cases completes MPI communication requests.
When the GPU control path used for starting and completion is the GPU stream scheduler, here and elsewhere this is termed \emph{stream-triggered communication} and generally does not involve GPU-callable routines; instead the host API interacts with the GPU stream control program to schedule implicit GPU 
starting
and completion of communication operations. In contrast, using GPU kernel thread execution for 
starting
and completion is generally termed \emph{kernel-triggered communication} and requires the addition of explicit GPU-callable APIs for 
starting
and completing communication.
As opposed to these options, 
 Intel's approach includes \emph{GPU-initiated APIs} (in which all parameters to the operation are provided in the GPU) and \emph{GPU-synchronization APIs} \cite{intel-mpi-reference,intel-mpi-gpu-buffers}.

Despite the many proposals, there is neither a consensus nor standard, either formal or informal, on an appropriate MPI API for GPU-triggered communication. To provide information and guidance for the MPI community to systematically address this issue (such as in MPI-5, or ideally sooner), this paper collects, organizes, and summarizes information about the various proposed MPI GPU-triggered communication approaches and highlights their similarities and differences. After providing background on representative GPU communication triggering proposals (Section~\ref{sec:example}), this paper makes  the following contributions:
\begin{itemize}
    \item Presents a taxonomy for approaches to GPU-triggered communication that highlights the major design decisions that an MPI API needs to address (Section~\ref{sec:taxonomy}); 
    \item Summarizes the key features of and classifies nine different MPI stream- and kernel-triggering API proposals using this taxonomy and compares and contrasts key requirements and API design elements of these approaches (Section~\ref{sec:classification});
    \item Highlights key gaps in the existing GPU triggering MPI APIs and in the MPI standard that will need to be addressed by future proposals for a robust MPI GPU triggering API (Section~\ref{sec:gaps}).
\end{itemize}
Finally, the paper offers conclusions and recommends a path forward in Section~\ref{sec:conclusion}.

\section{Representative GPU-Triggered Communication Abstractions}
\label{sec:example}

To illustrate the wide range of design choices and challenges in this area, we begin by providing information on four proposals for GPU-triggered MPI communication.
Specifically, we discuss the stream triggered, two-sided communication support in recent MPICH implementations~\cite{zhou_mpix_2022}, the stream-triggered MPI one-sided communication support recently proposed by researchers at Hewlett Packard Enterprise (HPE)~\cite{namashivayam_exploring_2023}, 
 extensions to the MPI 4.0 Partitioned Communication interface~\cite{mpi40} to support GPU kernel triggering of MPI communication~\cite{holmes_partitioned_2021},
 as well as 
 the GPU-initiated model proposed by Intel \cite{intel-mpi-reference,intel-mpi-gpu-buffers}.

We also present simple ping-pong communication for each proposal to further elucidate their proposed approaches. Together, this section highlights and provides essential background on the diverse design choices we explored in designing the GPU triggered communication taxonomy described below in Section~\ref{sec:taxonomy}. 

\subsection{MPICH Triggering}
The GPU triggering support in recent MPICH implementations \cite{zhou_mpix_2022} focuses on stream triggering of MPI communication using two abstractions:
\begin{description}
    \item[\texttt{MPIX\_Stream}]-- an MPI object to sequence MPI operations and to which external execution contexts, for example GPU streams, can be bound. This provides clearly defined concurrency semantics for MPI operations in multiple execution contexts including multi-threaded host execution contexts and accelerator-based MPI+X environments; and,
   { \item[GPU Stream communicators]-- MPI \hfill communicators \hfill to  \hfill which \hfill  an \hfill \texttt{MPIX\_-\\ Stream} object is attached to a GPU stream execution context, and on which MPI operations are enqueued to the attached GPU stream and return immediately in the host execution context.}
\end{description}
MPICH provides explicit enqueue variants of the supported GPU-triggered operations that return immediately to the host; GPU synchronization routines must be used to await for completion of these operations on the GPU. MPICH implementers considered making \verb|MPI_Send| and \verb|MPI_Recv| on communicators with GPU streams attached have enqueue semantics (similar to the previously proposed MPI-GDS API~\cite{venkatesh_mpi-gds_2017}), but in the end rejected this because it would change the semantics of well-known MPI APIs. Finally, MPICH stream communicators also include additional features to support APIs similar to the previous MPI Endpoints proposal~\cite{mpi-endpoints}.


\begin{figure}
\begin{lstlisting}[style=DerekStyle]
  MPIX_Stream mpistream;
  MPI_Comm stream_comm;
  MPI_Info info;
  
  MPI_Info_create(&info);
  MPI_Info_set(info, "type", "cudaStream_t");
  MPIX_Info_set_hex(info, "value", &cudastream, 
                    sizeof(cudastream));
  MPIX_Stream_create(info, &mpistream);
  MPI_Info_free(&info);
  MPIX_Stream_comm_create(MPI_COMM_WORLD, mpistream, 
                          &stream_comm);
  
  for (int i = 0; i < niters; i++) {
    if (my_rank == 0) {
      MPIX_Send_enqueue(src_buf, 1, MPI_INT, 1, 123, 
                        stream_comm);
      MPIX_Recv_enqueue(src_buf, 1, MPI_INT, 1, 123, 
                        stream_comm, MPI_STATUS_IGNORE);
    } else if (my_rank == 1) {
      MPIX_Recv_enqueue(dst_buf, 1, MPI_INT, 0, 123, 
                        stream_comm, MPI_STATUS_IGNORE);
      MPIX_Send_enqueue(dst_buf, 1, MPI_INT, 0, 123, 
                        stream_comm);
    }
  }
  cudaStreamSynchronize(cudastream);
\end{lstlisting}
\caption{Stream-triggered MPI Ping Pong using MPICH Stream Triggering\label{fig:mpich-triggering-pingpong} \cite{zhou_mpix_2022}}
\end{figure}

Figure~\ref{fig:mpich-triggering-pingpong} shows an example ping-pong communication using this interface. We highlight the following key features of this example:
\begin{itemize}
\item To use GPU stream-triggered communications, the programmer must create a new communicator;
\item The stream communicator concept, similar to the one previously proposed for MVAPICH~\cite{venkatesh_mpi-gds_2017}, can be naturally extended to include additional operations beyond the point-to-point operations supported in the current implementation, including collective communications; and,
\item MPICH's stream communicators enqueued operations require that the calling process must explicitly synchronize with the underlying GPU stream using GPU-specific operations to ensure completion of the corresponding MPI operation in the GPU execution context.
\end{itemize}

\subsection{HPE One-Sided}

The HPE one-sided GPU triggering API~\cite{namashivayam_exploring_2023} supports stream-triggered, one-sided communication\footnote{HPE also provides stream-triggered, two-sided communication, not considered here.}. 
%
The stream abstractions included in HPE's approach rely on GPU-NIC asynchronous stream triggering that integrates the NIC command queue and the GPU stream. In this approach, the CPU puts commands in the Cray Cassini NIC's command queue and then GPU stream operations triggers these commands. This removes unnecessary synchronization on the CPU. HPE's one-sided API is designed to allow for GPU-NIC progress where the host CPU is fully bypassed in progressing communication. HPE's previous two-sided GPU triggering APIs~\cite{namashivayam_exploring_2022} required host progress to receive messages because of the need for tag matching and unexpected message handling.

\begin{figure}
\vspace{-0.2in}
\begin{lstlisting}[style=DerekStyle]
  /* Normal MPI communicator, window, group, and  */
  /* buffers assumed to exist  */
  for(int i = 0; i < niters; i++) {
    if(rank == 0){
      /* Send ping */
      MPI_Win_start(group, MPI_MODE_STREAM, win);
      MPI_Put(src,n,MPI_INT,1, disp, n, MPI_INT,win);
      /* Puts triggered here */
      MPIX_Win_complete_stream(win,stream); 
      /* Receive pong */
      MPIX_Win_post_stream(group, win, stream);
      MPIX_Win_wait_stream(win,stream);
    } else { /* Receive ping */
      MPIX_Win_post_stream(group, win, stream);
      MPIX_Win_wait_stream(win, stream);
      /* Send pong */
      MPI_Win_start(group, MPI_MODE_STREAM, win);
      MPI_Put(src, n, MPI_INT, 0, disp, n, MPI_INT,win);
      /* Puts triggered here */
      MPIX_Win_complete_stream(win, stream);
    }
  }
  cudaStreamSynchronize(stream);
\end{lstlisting}
\caption{Stream-triggered MPI Ping Pong using HPE One-Sided Stream Triggering\label{HPE-triggering-pingpong}}
\end{figure}

%
%
%
Figure \ref{HPE-triggering-pingpong}, shows an example of a one-sided ping-pong test using the HPE Cray MPI interface where stream-triggered variants of window post, wait, and complete operations are combined with window start and put operations to achieve fully offloaded GPU communication.
We highlight the following key features of this example:
\begin{itemize}
    \item One-sided communication mixes new \verb|MPIX_*| functions and standard MPI functions, because the {\em Window} is associated with the stream.
    \item \verb|MPI_Put| and \verb|MPI_Win_start| do not have stream-triggered \verb|MPIX_| alternatives and simply initialize future communications operations to be triggered by the stream-triggered complete operation.
    \item The \verb|MPI_Win_complete_stream| \hfill starts \hfill communication \hfill operations \hfill in \hfill the \\ command queue initialized by the host (e.g., using \verb|MPI_Put|). This utilizes the GPU and NIC without weakening MPI RMA synchronization semantics, and in reality requires \emph{stronger} MPI RMA synchronization semantics by \emph{always} deferring one-sided operations until \verb|MPI_Win_complete_| \verb|stream| is executed by the GPU stream control program.  That is, in the HPE one-sided API, the communications indicated by \verb|MPI_Put| calls \emph{cannot} legally occur until \verb|MPI_Win_complete_stream| is executed in stream order by the GPU control program.
\end{itemize}

\subsection{Kernel-Triggered Partitioned Communication}

\begin{figure}
\begin{lstlisting}[style=DerekStyle]
__global__ void ready(MPIX_Prequest preq) {
  MPIX_Pready(threadIdx.x, preq);
}

__global__ void arrived(MPIX_Prequest preq) {
  int flag = -1;
  for(;;){
    MPIX_Parrived(preq, threadIdx.x, &flag);
    if(flag) break;
  }   
}

void pingpong(...) {
  int sendbuf[1], recvbuf[1];
  MPIX_Request srequest, rrequest;
  MPIX_Prequest psrequest, prrequest;
  int otherrank = (rank == 0) ? 1 : 0;
  MPIX_Psend_init(sendbuf, 1, 1, MPI_INT, otherrank, 0,
                  MPI_COMM_WORLD, MPI_INFO_NULL,
                  &srequest);
  MPIX_Precv_init(recvbuf, 1, 1, MPI_INT, otherrank, 0,
                  MPI_COMM_WORLD, MPI_INFO_NULL, 
                  &rrequest);
  MPIX_Prequest_create(srequest, &psrequest);
  MPIX_Prequest_create(rrequest, &prrequest);
  for(int i = 0; i < maxiters; i++) {
    MPIX_Start(&rrequest); // ping if rank==1, else pong
    MPIX_Start(&srequest); // pong if rank==1, else ping
    if(rank == 0) {
      ready<<<1, 1, 0, 0>>>(psrequest);
      MPIX_Wait(&srequest, MPI_STATUS_IGNORE);
      arrived<<<1, 1, 0, 0>>>(prrequest);
      MPIX_Wait(&rrequest, MPI_STATUS_IGNORE);
    } else {
      arrived<<<1, 1, 0, 0>>>(prrequest);
      MPIX_Wait(&rrequest, MPI_STATUS_IGNORE);
      ready<<<1, 1, 0, 0>>>(psrequest);
      MPIX_Wait(&srequest, MPI_STATUS_IGNORE);
    }
  }
}

\end{lstlisting}
\caption{
MPI Ping Pong using MPI-ACX Partitioned Stream Triggering\label{mpiacx-triggering-pingpong}}
\end{figure}

The MPI-ACX implementation \cite{dinan_james_mpi-acx_2023} includes a prototype of the partitioned communication detailed in MPI-4.
Partitioned point-to-point communication was specifically added to MPI-4 for a variety of reasons, including providing matching semantics (matching at initialization) and lightweight interfaces for data transfer initiation\footnote{It also removes the restrictions of persistent send and receive operations---\texttt{MPI\_Send\_init} and \texttt{MPI\_Recv\_init}---from MPI-1, which are forbidden to communicate and do not match only once.  These fail to offer a channel semantic or obviate receive-queue matching. By way of contrast, single partitioned mode provides a uni-directional channel abstraction.}.  A specific nuance in this approach is that both \verb|MPIX_| \verb|Parrived| and \verb|MPIX_Pready| can be called from the host and the CUDA device.  There are device-variants of the host requests to support the device-invoked operations.

Notably, the completion of an \verb|MPI_Psend_init| call does not guarantee that a receiver has been matched
prior to calling  \verb|MPI_Start|; 
it can instead occur at the first \verb|MPI_Wait|
for both sides of the partitioned channel.  As such, proposals were offered for MPI-4.1 denoted \texttt{MPI\_Pbuf\_\-prepare} to guarantee initialization completion prior to first use and ensure readiness of buffers upon reuse \cite{forum_synchronization_2020}. This has not yet been adopted for the standard, but appears essential to avoid unexpected messages that would introduce complex logic that is considered untenable in GPU kernels.

Figure \ref{mpiacx-triggering-pingpong} details a ping-pong test based on the ring program supplied with MPI-ACX;  and the following are some key observations:
\begin{itemize}
    \item In \verb|MPI-ACX|, the programmer uses \verb|MPIX_Request| to create a \verb|MPIX_Prequest| for the device to use in operations. These objects are tied to each other with \verb|MPIX_Prequest|\verb|_create|.
    \item The host-based MPI request object is used for regular MPI functions such as \verb|MPIX_Start| and \verb|MPIX_Wait|.
    \item Different request types and function signatures are needed for GPU and CPU interfaces due to differences between the state needed in each context.
\end{itemize}

\subsection{Intel GPU-Initiated Communication}
Intel provides a GPU-initiated communication feature \cite{intel-mpi-gpu-buffers} that appears similar to HPE's one-sided communication prototype.
Intel's additions to its MPI implementation \cite{intel-mpi-reference} is kernel-triggered, and uses existing SYCL queues for communication ordering instead of an explicit MPI ordering object (as MPICH does).
Figure~\ref{fig:intel-GPU-initiated} demonstrates the ping-pong exemplar following Intel's approach that reduces the number of MPI functions to only \verb|MPIX_Put_notify| and \verb|MPIX_-| \verb|Win_get_notify| instead putting work on the GPU using a SYCL queue object and adding them as events\footnote{Personal Communication, Dr.\@ Daniel J.\hbox{} Holmes, Intel Corp., June 14, 2024}. 

\begin{figure}

\begin{lstlisting}[style=DerekStyle]
void do_pingpong_mpi(int target, sycl::queue &q, [[maybe_unused]] void **ptrs, int threads=1) {
    double latency = 0.0;   
    int tgt = target;
    MPI_Win_fence(0, win);
    for (int i=0; i<2;++i) {
        int iter = iters[i];
        if (rank == 0) {
            sycl::event e = q.submit([&](sycl::handler &h) {
                h.parallel_for_work_group(sycl::range(1), sycl::range(threads), [=](sycl::group<1> grp) {
                    grp.parallel_for_work_item([&](sycl::h_item<1> item) {
                        int id = item.get_global_id(0);
                        int v = 0;  MPI_Count c = 0;
                        for (int i=1; i<iter; ++i) {
                            MPIX_Put_notify(&v, 1, MPI_INT, tgt, id, 1, MPI_INT, id, win);
                            while (c<i)
                                MPIX_Win_get_notify(win, id, &c);
                        }    });    });    });
            q.wait();
        } else if (rank == target) {
            q.submit([&](sycl::handler &h) {
                h.parallel_for_work_group(sycl::range(1), sycl::range(threads), [=](sycl::group<1> grp) {
                    grp.parallel_for_work_item([&](sycl::h_item<1> item) {
                        int id = item.get_global_id(0);
                        int v = 0; MPI_Count c = 0;
                        for (int i=1; i<iter; ++i) {
                            while (c<i)
                                MPIX_Win_get_notify(win, id, &c);
                            MPIX_Put_notify(&v, 1, MPI_INT, 0, id, 1, MPI_INT, id, win);
                        }    });    });    }).wait();
        }   }
    MPI_Win_fence(0, win);
    for (i = 0; i < threads; ++i)
        MPIX_Win_set_notify(win, i, (MPI_Count)0);
}
\end{lstlisting}
\caption{Ping Pong using Intel GPU-Initiated MPI \label{fig:intel-GPU-initiated}\cite{intel-mpi-reference,intel-mpi-gpu-buffers}}
\end{figure}

\section{A Taxonomy of MPI GPU-Triggered Communication}
\label{sec:taxonomy}

Based on a careful examination of both the MPI standard and prior proposals for GPU-triggered communication, we defined the following categories for classifying key characteristics of MPI GPU-triggered communication proposals in four different overarching areas:

\begin{description}
    \item[\emph{Area 1: GPU control path used:}] Does the API support the GPU stream control path or kernel control path for 
    starting
    and completing MPI communication operations? In the case of APIs where both stream- and kernel-triggering are supported, we split the API into separate stream- and kernel-triggered sub-APIs and classify them separately because of the frequent semantic and API differences between the such approaches in other categories.

    \vspace{0.1in}
    \item[\emph{Area 2: API Design Considerations}]
    \item[Reuses Existing MPI APIs or abstractions:] \hfill Does \hfill the \hfill GPU-triggering \\ API reuse existing MPI communication operations or abstractions to enable GPU triggering of communication operation starting or completion, or does it instead create new MPI operations or abstractions?
    \item[Changes Existing MPI API Semantics:] If the API reuses existing APIs or abstractions, does it change the semantics of these API elements to behave differently than they currently do? For proposals that modify the API, we classify whether the API \emph{strengthens} MPI API semantics in a way that preserves backward compatibility or \emph{weakens} MPI semantics in a way that may compromise backward compatibility.
    \item[Separate MPI Operation Initialization and Starting:] Does the API use a single call to execute the \emph{initiate} stages of MPI communication operations or does it separate this into explicit API calls to \emph{initialize} and \emph{start} MPI communication operations?
    \item[GPU MPI Operation Completion Support:] Does the API support GPU detection of the completion of GPU-initiated MPI operations? We identify the classifications of \emph{all}, \emph{some}, and \emph{none} for this category, corresponding to GPU control paths supporting waiting or polling for the completion of (1) all GPU-initiated MPI operation, (2) some GPU-initiated MPI operations, or (3) requiring host code to wait or poll for the completion of MPI operations. 
    \item[Collective Communication Support:] Does the API specify operations enabling a single GPU communication operation to communicate with multiple processes? We classify collective support as either \emph{full}, \emph{partial}, \emph{group}, or \emph{none}, corresponding to (1) an API that supports the full semantics of MPI two-sided collective operations, (2) an API that supports a subset of the MPI two-sided collective operations,  (3) an API that supports collective pairwise exchanges between subgroups of a communicator such as one-sided group or two-sided neighbor collectives, or (4) no specified support for collective communication.

    \vspace{0.1in}
    \item[\emph{Area 3: Ordering and Concurrency Considerations}]
    \item[MPI Operation Sequencing Abstraction:] Does the API include an explicit MPI abstraction or object for ordering GPU communication operations separate from GPU-provided abstractions?
    \item[Sequencing Abstraction Semantics:] If the API includes an explicit sequencing abstraction, are communication operations sequenced using
    it fully ordered or partially ordered (e.g.,  using a fencing or graph abstraction)?
    \item[Specification of MPI Concurrency Standard Integration:] Does the API address concurrency-related issues arising from GPU triggering of MPI communication through \emph{explicit} specification or \emph{implicit} specification via the existing MPI \emph{thread multiple} mode, or are concurrency semantics \emph{unspecified}?

    \vspace{0.1in}
    \item[\emph{Area 4: Implementation Considerations}]
    \item[GPU/NIC Progress:] Does the API enable message initiation or starting and completion by the GPU and network device without the involvement of the host CPUs? This is generally accomplished by supporting one-sided data movement that avoids the control flow complexity of matching in the receive path.
    \item[Available Implementation or Evaluation:] Is an implementation or evaluatio of this API available either for download and installation, from a vendor, or in a peer-reviewed publication so that it can potentially be tested or compared with other APIs? 
    \item[Multi-architecture Support:] Does the implementation, if available, target a single vendor architecture or has it been demonstrated for multiple GPU and/or NIC accelerator architectures? We categorize an implementation as \emph{Yes} when it supports multiple GPUs and network architectures, as \emph{GPU} or \emph{NIC} when it has been demonstrated on multiple GPU or NIC architectures, respectively, and as \emph{No} when it has been demonstrated on neither multiple NIC nor multiple GPU architectures. 
\end{description}

\section{Classifying Existing MPI GPU Triggering Proposals}
\label{sec:classification}

In this section, we classify these proposals and their availablity or described implementations according to the taxonomy described in Section~\ref{sec:taxonomy}, as shown in Table~\ref{tab:taxonomy}. We then use this taxonomical classification to compare and contrast key features of these proposals and their corresponding implementations.

\subsection{API Classification} 
We focus on the nine proposed GPU-triggered MPI programming interface extensions listed below: 
\begin{description}
    \item[MPI-GDS~\cite{venkatesh_mpi-gds_2017}] -- a GPU-triggering interface extension for the MVAPICH MPI implementation~\cite{MVAPICH}.
    \item[MCI-ACX Enqueued~\cite{dinan_james_mpi-acx_2023}] -- stream-triggered, two-sided communication operations as implemented in the MPI-ACX GPU-triggering interface extension for NVIDIA network devices that support MLX4 device functionality.
    \item[MPICH Triggering~\cite{zhou_mpix_2022}] -- a GPU-triggering interface \hfill extension \hfill for \hfill the \\ MPICH MPI implementation~\cite{MPICH2}.
    \item[HPE Send/Recv~\cite{namashivayam_exploring_2022}] -- a stream-triggered MPI interface extension for HPE systems with the Slingshot network interface~\cite{hpe-slingshot}.
    \item[Project Delorean~\cite{koziol_composable_2023}] -- a proposal for graph-based sequencing of MPI operations that including GPU triggering.
    \item[HPE One-sided~\cite{namashivayam_exploring_2023}] -- a GPU-triggered Post/Start/Complete/Wait interface extension for HPE systems with the Slingshot network interface.
    \item[Partitioned Communication] -- the MPI partitioned communication API~\cite{mpi40} supplemented with the \verb|Pbuf_prepare| and partitioned collective operations recently proposed~\cite{holmes_partitioned_2021}, and partially implemented by MPI-ACX~\cite{dinan_james_mpi-acx_2023}.
    \item[HPE Persistent~\cite{hewlett-packard_enterprise_mpix_start_2024}] -- a kernel-triggered communication MPI interface extension for HPE systems with the Slingshot network interface~\cite{hpe-slingshot}.
    \item[Intel GPU-Initiated~\cite{intel-mpi-reference,intel-mpi-gpu-buffers}] -- a  
    GPU-initiated extension for Intel MPI\footnote{Personal Communication, Dr.\@ Daniel J.\hbox{} Holmes, Intel Corp., June 14, 2024}.
\end{description}
Note that while multiple variants of some of these designs 
have been proposed in the MPI Forum Working groups, BOFs, and other venues, we focus on the above listed API descriptions and implementations as representative of the main existing approaches at the time of this publication.
\begin{table}[p]
\begin{center}
\begin{threeparttable}
\caption{Taxonomy
of MPI GPU-Triggered Communication Proposals
 \label{tab:taxonomy}}
\begin{tabular}{|l|l|l|l|l|l|l|}
\hline
        & Area 1 &        \multicolumn{5}{c|}{Area 2: API}                                       \\
        &        &        \multicolumn{5}{c|}{Features}                                  \\
\hline
Proposal& Control& Reuses  & Changes    & Separate    & GPU        & Collective\\ 
        & Path   & Existing& API        & Initialize  & Completion & Support   \\
        & Used   & APIs    & Semantics  & and Start& Support    &           \\
\hline\hline
MPI-GDS &  Stream& Yes     & Weaker    & No          & Full        & Full\tnote{1} \\
\hline
MPI-ACX & Stream & Yes\tnote{1}& No     & Yes\tnote{1}& Full        & No \\
Enqueued & &&&&&\\
\hline
MPICH   & Stream  & Yes    & No\tnote{2}& No          & Full        & Partial    \\
Triggering & &&&&&\\
\hline
HPE     & Stream  & No     & No         & Yes         & Full        & No         \\ 
Send-Recv & &&&&&\\
\hline
Delorean& Stream  & No\tnote{3}& No    & Yes         & Full        & Full        \\
\hline
HPE     & Stream  & Yes     & Stronger\tnote{4} & Yes& Full        & Group   \\
One-sided & &&&&&\\
\hline
Partitioned& Kernel& Yes   & No         & Yes         & Partial\tnote{5}     & Full\tnote{1} \\
Comm. &&&&&&\\
\hline
HPE     & Kernel    & Yes  & No         & Yes         & Full         & No         \\
Persistent &&&&&&\\
\hline
Intel     & Kernel    & Yes  & No         & No         & Full         & No         \\
GPU-Init &&&&&&\\
\hline
\end{tabular}

\medskip

\begin{tabular}{|l|l|l|l|l|l|l|}
\hline
        & \multicolumn{3}{c|}{Area 3: Ordering and}   & \multicolumn{3}{c|}{Area 4: Implementation} \\
        & \multicolumn{3}{c|}{Concurrency}    & \multicolumn{3}{c|}{Considerations} \\
\hline
Proposal& MPI        & Sequencing& Concurrency& GPU/NIC  & Impl.  & {Multiple}\\ 
        & Sequencing & Model     & Model      & Progress & Available & {Arch.} \\
        & Object     & Semantics & Specified  &          &    & {Support} \\
\hline\hline
MPI-GDS & No         & N/A       & Implicit   & No       & Yes       & No \\
\hline
MPI-ACX & No         & N/A       & No         & No       & Yes       & NIC \\
Enqueued &&&&&&\\
\hline
MPICH   & Yes        & Full      & Explicit   & No       & Yes       & Yes\\
Triggering &&&&&&\\
\hline
HPE     & Yes        & Partial   & No         & No       & Yes       & GPU \\ 
Send/Recv &&&&&&\\
\hline
Delorean& Yes        & Partial   & Implicit   & No       & No        & No \\
\hline
HPE     & No         & N/A       & Implicit   & Yes      & Yes       & GPU \\
One-sided  &&&&&&\\
\hline
Partitioned & No     & N/A       & Explicit   & Yes      & Yes       & No \\
Comm. &&&&&&\\
\hline
HPE     & No         & N/A       & No         & No       & Yes       & GPU\\
Persistent &&&&&&\\
\hline
Intel     & No       & N/A       & Yes         & Yes     & Yes       & GPU\\
GPU-Init &&&&&&\\
\hline
\end{tabular}
\begin{tablenotes}
\item[1] Proposed in publication but not in the described or available implementations.
\item[2] The initial proposal considered changing the semantics of \verb|MPI_Send| and \verb|MPI_Recv| on GPU stream communicators to be identical to their \verb|_enqueue| variants and return to the caller prior to local completion of the operation; MPICH developers have since abandoned this approach.
\item[3] Proposes ``deferred'' operations as a generalization of MPI persistent operations.
\item[4] Explicitly defers \verb|MPI_Put|/\verb|MPI_Get| execution to when \verb|MPI_Win_complete_stream| is called.
\item[5] There is no GPU call to determine if a partition that has been marked ready to send has completed locally.
\end{tablenotes}
\end{threeparttable}
\end{center}
\end{table}

\subsection{API Feature Comparisons}
Based on the comparison shown in Table~\ref{tab:taxonomy}, there is no clear agreement on whether stream or kernel triggering is preferable, either from an API, programmability/usability, implementation, or performance perspective.
Most implementations seek to reuse portions of the API and maintain or strengthen the semantics of the API operations they use, and also separate initialization and initiation of API operations. 

Almost every API allows the GPU to check for or wait for the completion of every MPI operation that they either initiate or start, enabling the GPU to avoid returning to the host execution context to ensure communication completion. The main exception is partitioned communication, where receives completion can be tested on the GPU using \verb|Parrived|, but there is no equivalent call for testing for the completion of sends initiated from the kernel using \verb|Pready| (the only available option is to check if all send partitions have completed using a single call to \verb|Wait| or \verb|Test| or one of their variants).

While several APIs propose collective support, only limited support for any GPU-triggered collectives is available, e.g., \verb|MPIX_Allreduce_enqueue| in MPICH and point-to-point group communication in the HPE one-sided triggering API. This is likely due to the complexity of combining complex collective communication with the constrained GPU and NIC control paths. 

There is little agreement on whether to introduce a new MPI-level primitive to represent GPU execution ordering, with some APIs introducing new primitives and others reusing opaque GPU streams. Note that this is the case even from the same vendor, where the  HPE two-sided API includes an \verb|MPI_Queue| object that is not used by the HPE one-sided API.

Implementation availability  varies significantly. For instance, Delorean is just a specification with no known implementation; Intel's GPU support is currently in ``preview,'' while MPI-ACX is fully available and so we were able to run the ping-ping example on representative hardware. 
Few implementations support full GPU/NIC offload of communication; most APIs designed with the assumption that the host will be involved in data movement. 

\section{Key Gaps}
\label{sec:gaps}

Based on the classifications described in the previous section, this section  covers gaps and perceived limitations observed in prior work and prototypes as well as the MPI standard.

\subsection{Limitations in the MPI Standard}
In the MPI Standard, there are two main areas that needs to be addressed.
\begin{enumerate}
\item Problems surrounding persistence:
The \hfill separation \hfill of \hfill initialization \hfill  and \hfill starting \hfill in \hfill GPU \hfill triggering \hfill APIs \hfill make \hfill persistent \hfill operations \hfill attractive \hfill for \hfill this \hfill purpose, \hfill as \hfill demonstrated \hfill by \hfill partitioned \hfill communication. \hfill Unfortunately,  \hfill MPI-1 specified \hfill persistent \hfill send \hfill and \hfill receive \hfill operations \hfill cannot \hfill communicate \hfill at \hfill initialization \hfill  nor \hfill does \hfill their \hfill initialization  \hfill  result \hfill in \hfill matching; \hfill indeed, persistent and non-persistent operations can match each time they are started.



%

\item Lack of \texttt{MPI\_Pbuf\_prepare*} 
operations for two-sided communication:
Two separate issues that have been conflated in  prior discussions of \texttt{MPI\_Pbuf\-\_prepare} and \texttt{MPI\_Pbuf\_prepareall}, 
yet to be standardized \cite{forum_synchronization_2020}:
\begin{enumerate}
\item     Ensuring that MPI two-sided requests are matched (once and for all) so that
there is agreement between the sender and receiver side about the exposed buffer on which to make operations (such as with \texttt{MPI\_Put}s for early-bird communication as \texttt{MPI\_Pready}'s are issued on the send side).
\item     Ensuring or asserting that the remote buffer is \emph{ready} for one-sided data movement; that is, there are no unexpected transfers.
\end{enumerate}
The former evidently subsumes the latter; otherwise, matching/synchronization would have to be done prior to every call to \texttt{MPI\_Start}, which is likely to be unacceptably costly.
%
Further, because of the allowance for weak progress in MPI implementations, there is no guarantee even in two-sided operations that match at initialization (e.g., in partitioned communication) that the initial match has occurred between 
before the first complete communication. 
Thus, initial communication can occur before any agreement on the readiness of the receive buffer 
(e.g., agreement and transmission could first transpire during the first wait calls, and/or the initial transfers could be orchestrated differently than subsequent ones).


In addition, the desire for these conditions is not unique to partitioned \hfill communication---the various \hfill stream-triggered \hfill communication \hfill proposals \\ (MPICH, HPE two-sided, MPI-ACX enqueuing, etc.) would want this functionality to push their data movement through the one-sided path as well. 
\end{enumerate}


\subsection{Limitations across Proposed APIs}
In addition to gaps in the MPI standard that increase the difficulty of supporting GPU triggering, there are also recurring limitations across the various APIs that have been proposed. We highlight these gaps below to encourage efforts to address them consistently in new and revised proposals:
\begin{itemize}
\item Clear concurrency model---If the host execution  context and accelerator context both initiate or start operations, the ordering and concurrency is ill-defined in many APIs, are left to the vagueness of MPI, or are silent. 
MPICH stream triggering is the only one that provides a clear, specific concurrency model for both host- and accelerator-initiated communication.
\item Matching---There is no explicit support for matching of send/receive buffers so that this occurs only once, and receive matching overheads are eliminated.
\item GPU completion support with partitioning---If something is  started  on a GPU, there is no way to know if something is completed on the send side. This makes it difficult or impossible to write programs without need to synchronize with the CPU to interleave communication and computation.
\item Collective communication support---This is broadly absent yet needed to support real applications that avoid unnecessary synchronization with the CPU. In particular, \emph{no} current MPI stream triggering approach supports stream triggering of the full range of MPI collective communication operations.
\item GPU/NIC Progress---Lack of this feature is a gap for any performant implementation. Out of the nine APIs considered only three (partitioned communication augmented with \verb|Pbuf_prepare|, HPE one-sided communication, and Intel GPU-initiated) support GPU/NIC progress. 
\end{itemize}

\section{Conclusion}
\label{sec:conclusion}


Message-passing communication involving GPUs with MPI has proven to be both complex and expensive, motivating new approaches. 
This is the first comprehensive survey of all nine proposals known  to the authors that address stream/\\graph-, \hfill kernel-triggered, \hfill and \hfill GPU-initiated \hfill communication \hfill abstractions \hfill for \\ MPI, whose principal purpose is to improve the performance of communication when GPU kernels create or consume data for transfer through MPI operations.
%

Here, we compared and contrasted stream/graph- and kernel-triggered MPI communication abstractions.  Various proposals with distinct  potential APIs for stream- and/or kernel-triggering have been proffered as designs and/or prototypes, which span various GPU architectures and approaches, including MPI-4 partitioned point-to-point communication, stream communicators,  explicit MPI stream/queue objects, and notified-RMA. 
%
Notably, some of these proposals either strengthen or weaken the semantics of MPI operations.
We enumerated nine different proposals and provided a taxonomy of these  proposals and their operations against twelve categories of features and semantics.

We described the design space in which these abstractions reside, their 
use of stream and other non-MPI abstractions, their relationship to partitioned and persistent operations, and discussed their potential for added performance, how usable 
these are, and where 
functional and/or semantic 
gaps exist.  
%
Designs breaking backward compatibility with MPI were also duly noted (and cautioned).
We also presented code examples of how some of these operate on the ubiquitous ping-pong use case (variously utilizing  CUDA,   ROCM, and Intel  APIs).

The next steps to advance performance-portable programming with acceler\-ator-based MPI+X programming should start with robust, inclusive conversations and design bake-offs among the various proposers, implementers, and users to identify a coherent path forward within the MPI community, inclusive of  application developers.  One or more proposed standard APIs 
would be competed then down-selected for MPI-5.
Existing proposals/prototypes cited in this paper may not include all hardware vendors or systems currently available and certainly there may be other solutions forthcoming or that are not yet public (e.g., for FPGAs  or other GPU accelerator products). 
Yet, fitting additional
 and/or forthcoming proposals, prototypes, and designs into our  taxonomy will provide a coherent basis to compare and contrast these with the nine presented here.

\begin{credits}
\subsubsection{\ackname} 

The authors wish to acknowledge Dr.\hbox{} Daniel J.~Holmes of Intel Corporation for providing information, examples, and references on the Intel GPU triggering API for this paper. The authors also wish to acknowledge Dr.\hbox{} Hui Zhou of Argonne National Laboratories for providing additional information and clarifications on the MPICH stream triggering API.

This work was performed with partial support from the National Science
Foundation  under Grants Nos.~ 
OAC-2103510, CCF-2405142 and CCF-2412182,
the U.S.\hbox{} Department of Energy's National Nuclear Security Administration (NNSA) under the Predictive Science Academic Alliance Program (PSAAP-III), Award DE-NA0003966, and Tennessee Technological University.
%
%
%
Any opinions, findings, and conclusions or recommendations expressed in this material are those of the authors and do not necessarily reflect the views of the 
National Science Foundation, or the U.S.\hbox{} Department of Energy's National Nuclear Security Administration.

\subsubsection{\discintname}
The authors have no competing interests to declare that are
relevant to the content of this article.
\end{credits}
%
%
%
 \bibliographystyle{splncs04}
 \bibliography{references,zotero}

\end{document}